\documentclass{emulateapj}

\usepackage{color}
\usepackage{times}
\usepackage{fancyhdr}
\usepackage{float}
\usepackage{alltt}
\usepackage{listings}
\usepackage{enumerate}
\usepackage[T1]{fontenc}
\usepackage{ae}
\usepackage{rotating}
\usepackage{subfigure}
\usepackage{amsmath}
\usepackage{amssymb}
\usepackage{graphicx}
\usepackage{dcolumn}
\usepackage{bm}
\usepackage{aas_macros_pk}

\def\be{\begin{equation}}
\def\ee{\end{equation}}
\def\beg{\begin{align}}
\def\eeg{\end{align}}
\def\bi{\begin{itemize}}
\def\ei{\end{itemize}}
\def\ben{\begin{enumerate}[1.]}
\def\een{\end{enumerate}}

\newcommand{\abs}[1]{\left| {#1} \right|}

\newcommand{\bo}{\raise-1mm\hbox{\Large$\Box$}}
\newcommand{\sci}[2]{#1 \times 10^{#2}}
\newcommand{\rthz}{\mathrm{Hz}^{-\frac{1}{2}}}
\newcommand{\hrss}{h_{\mathrm{rss}}}

\newcommand{\hrssn}{h_{\mathrm{rss}}^{90\%}}

\newcommand{\egwn}{E_{\mathrm{GW}}^{90\%}}
\newcommand{\egw}{E_{\mathrm{GW}}}
\newcommand{\eem}{E_{\mathrm{EM}}}
\newcommand{\gamman}{\gamma_{\mathrm{UL}}}

\newcommand\ligodoc{P0900024}

\keywords{gravitational waves - soft gamma repeaters}

\shorttitle{Stacked Search for Gravitational Waves from the 2006 SGR 1900+14 Storm}
\shortauthors{The LIGO Scientific Collaboration}

\begin{document}

\submitted{April 30, 2009}
\title{Stacked Search for Gravitational Waves from the 2006 SGR 1900+14 Storm}

\author{B.~P.~Abbott\altaffilmark{17},
 R.~Abbott\altaffilmark{17},
 R.~Adhikari\altaffilmark{17},
 P.~Ajith\altaffilmark{2},
 B.~Allen\altaffilmark{2, 60},
 G.~Allen\altaffilmark{35},
 R.~S.~Amin\altaffilmark{21},
 S.~B.~Anderson\altaffilmark{17},
 W.~G.~Anderson\altaffilmark{60},
 M.~A.~Arain\altaffilmark{47}
% ...
 }

\affil{The LIGO Scientific Collaboration, http://www.ligo.org}

% ...
\altaffiltext {2}{Albert-Einstein-Institut, Max-Planck-Institut f\"{u}r Gravitationsphysik, D-30167 Hannover, Germany}

% ...
\altaffiltext {17}{LIGO - California Institute of Technology, Pasadena, CA  91125, USA}

% ...
\altaffiltext {21}{Louisiana State University, Baton Rouge, LA 70803, USA}

% ...
\altaffiltext {35}{Stanford University, Stanford, CA  94305, USA}

% ...
\altaffiltext {47}{University of Florida, Gainesville, FL  32611, USA}

% ...
\altaffiltext {60}{University of Wisconsin-Milwaukee, Milwaukee, WI 53201, USA}

% ...

\begin{abstract}

We present the results of a LIGO search for short-duration gravitational waves (GWs) associated with the 2006 March 29 SGR 1900+14 storm.  
A new search method is used, ``stacking'' the GW data around the times of individual soft-gamma bursts in the storm to enhance sensitivity for models in which multiple
bursts are accompanied by GW emission.  We 
assume that variation in the  time difference between burst electromagnetic emission and potential burst GW emission is small relative to the  
GW signal duration, and we time-align GW excess power time-frequency tilings containing individual burst triggers to their corresponding 
electromagnetic emissions.    We use two GW emission models in our search:  a fluence-weighted model and a flat (unweighted) model for the most electromagnetically energetic bursts.  
We find no evidence of GWs associated with either model.   Model-dependent GW strain, isotropic GW emission energy $\egw$, and $\gamma 
\equiv \egw / \eem$ upper limits  are estimated using a variety of assumed waveforms.  The stacking method allows us to set the most stringent model-dependent limits on transient GW strain published to date.     We find $\egw$ upper limit estimates (at a nominal distance of 10\,kpc) of between $\sci{2}{45}$\,erg and $\sci{6}{50}$\,erg depending on waveform type.  These limits are an order of magnitude lower than upper limits published previously for this storm and overlap with the range of electromagnetic energies emitted in SGR giant flares.

\end{abstract}


\maketitle

\section{Introduction}

\begin{figure*}[!t]
\subfigure{
\includegraphics[angle=0,width=180mm,clip=false]{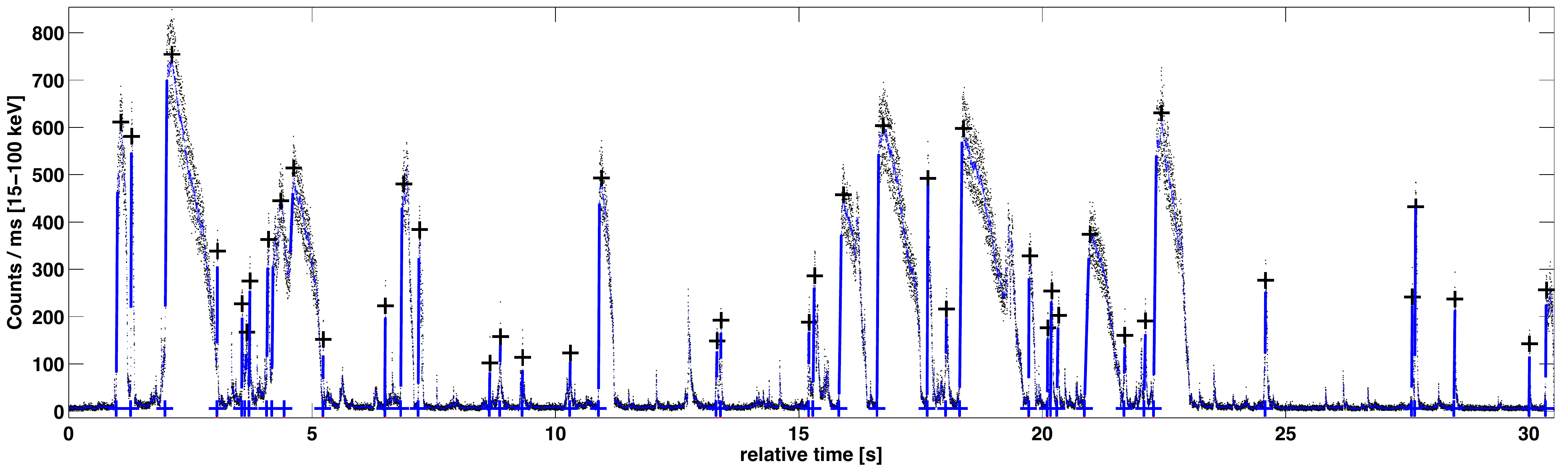}}
\subfigure{
\includegraphics[angle=0,width=180mm,clip=false]{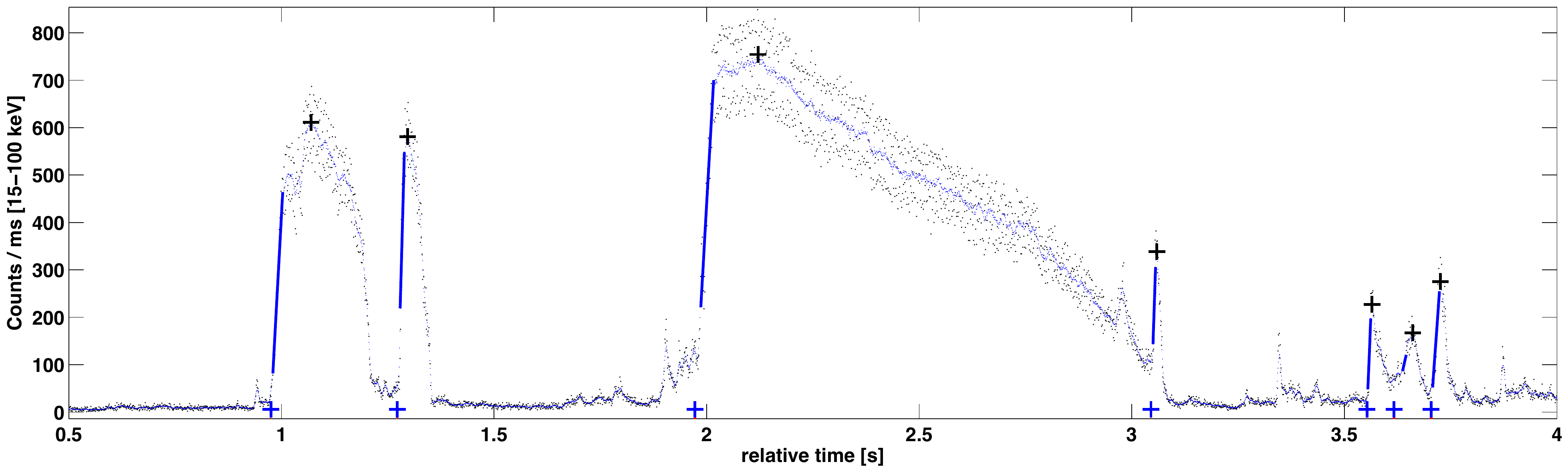}}
\caption[] { SGR 1900+14 storm light curve with 1\,ms bins in the (15--100)\,keV band.  Bottom plot shows a detail.  Burst start times are estimated by fitting the steeply rising burst edges; 
EM fluences are estimated  by integrating light curve area under each burst.   A 30-bin running average is shown in addition to the raw light curve.  Solid lines are linear fits to rising 
edges; the boundaries of rising edges were found by examining the first derivatives in the neighborhoods of the peak locations.  Crosses mark burst peaks and intersections of the rising edge fits extrapolated to a linear fit of the noise floor measured in a quiescent period of data in the 50\,s BAT sequence before the start of the storm.  The one-sigma timing uncertainty averaged over all measurements is $2.9$\,ms.    X-axis times are relative to 2006-03-29 02:53:09.9 UT at the Swift satellite.}
\label{fig:lc_rises}
\end{figure*}

Soft gamma repeaters (SGRs) sporadically emit brief 
($\sim0.1$\,s) intense bursts of soft gamma-rays. Three of the five known SGRs have produced rare ``giant flare'' events with 
initial bright, short ($\sim0.2$\,s) pulses with peak electromagnetic (EM) luminosities between $10^{44}$ and $10^{47}$\,erg~s$^{-1}$,
placing them among the most EM luminous events in the Universe.   According to the ``magnetar'' model SGRs are galactic neutron 
stars with extreme magnetic fields $\sim\nolinebreak10^{15}$\,G\,\citep{duncan92}.  Bursts  may result from the interaction of the star's 
magnetic field with its solid crust, leading to crustal deformations and occasional catastrophic cracking\,\citep{thompson95, schwartz05, horowitz09} with 
subsequent excitation of nonradial neutron star $f$-modes\,\citep{andersson97, pacheco98, ioka01} and the emission of GWs\,\citep{owen05, horvath05, 
pacheco98, ioka01}.   For reviews, see \,\cite{mereghetti08, woods04a}.

Occasionally SGRs produce many soft-gamma bursts in a brief period of time; such intense emissions are referred to as ``storms.''  We present a search for short-duration GW signals ($\lesssim$0.3\,s) associated with \emph{multiple} bursts in the 2006 March 29 SGR 
1900+14 storm\,\citep{israel08} using data collected by the Laser Interferometer Gravitational Wave Observatory (LIGO) \,\citep{S5}.   The storm light curve, obtained from the Burst Alert Telescope (BAT) aboard the Swift satellite\,\citep{BAT},  is shown in Fig.\,\ref{fig:lc_rises}.   It consists of more than 40 bursts in $\sim$30\,s, including common SGR bursts and some intermediate flares with durations $>0.5$\,s.  The total fluence for the storm event was estimated by the Konus-Wind team to be $\sci{(1-2)}{-4}$\,erg\,cm$^{-2}$ in the (20--200)\,keV range\,\citep{gcn4946}, implying an isotropic EM energy  $\eem = \sci{(1-2)}{42}$\,erg at a nominal distance to SGR 1900+14 of 10\,kpc (source
location and distance is discussed in\,\cite{kaplan02b}).  At the time of the storm both of the 4\,km LIGO detectors (located at Hanford, WA and Livingston, LA) were taking science quality data.  

\begin{figure*}[!t]
\begin{center}
\includegraphics[angle=0,width=180mm, clip=false]{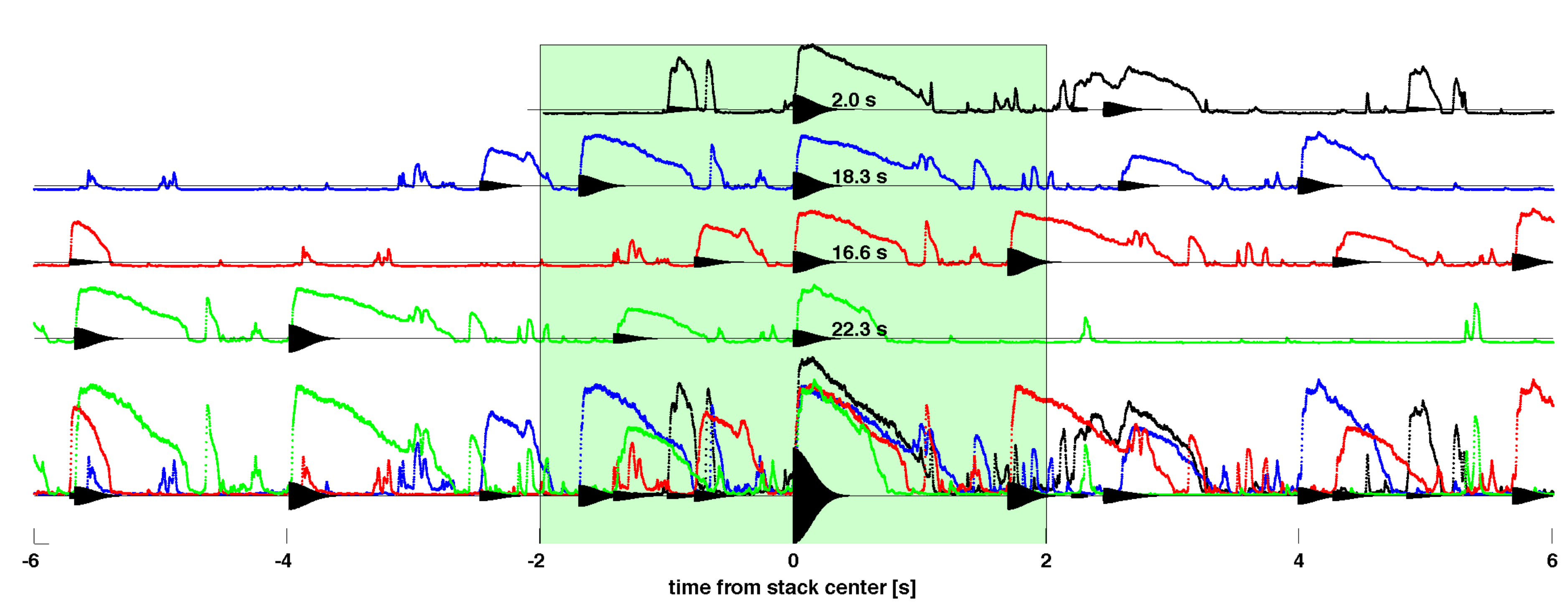}
\caption[] {Individual EM bursts inform the stacking of GW data.  This figure suggests the stacking procedure and explicitly shows search timescales.  The top four plots are 
EM light curve time series around individual bursts beginning at 2.0 s, 16.6 s, 18.3 s, and 22.3 s in Fig. \,\ref{fig:lc_rises}.  Simulated GW ringdowns in the fluence-weighted model are superposed.    The bottom plot shows the EM time series simultaneously, and the sum of the hypothetical GW signals.
The on-source region of $\pm2$\,s is shaded.    In the search, GW data corresponding to the EM time series are transformed to time-frequency power tilings before being added together and therefore there is no dependence on phase-coherence of GW signals in the analysis; this transformation is not illustrated.  } \label{fig:lc_stack}
\end{center}
\end{figure*}

We attempt to improve sensitivity to multiple weak GW burst signals associated with the storm's multiple EM bursts by adding together GW signal power over multiple bursts.   In doing so we assume particular GW emission models, which we describe in the next section.  Fig. \,\ref{fig:lc_stack} illustrates the stacking procedure using  the four most energetic bursts in the storm.

\section{Methods} 

The analysis is performed by the Stack-a-flare pipeline\,\citep{stackMethod}, which extends the method used in a recent LIGO  search for transient GW associated with individual SGR bursts\,\citep{s5y1sgr} and relies on an excess power detection statistic~\citep{anderson01}.   To ``stack'' $N$ bursts in the storm, we first generate $N$ excess power time-frequency
tilings.  These are  2-dimensional matrices in time and frequency generated from the two detectors' data streams.  Each tiling element gives an excess power estimate in the GW detector data stream in a small period of time $\delta t$ and a small range of frequency $\delta f$.  The time range of each tiling is chosen to be centered on the time of one of the target EM bursts in the storm.  We then align these $N$ tilings along the time dimension so that times of the target EM bursts coincide, and perform a weighted addition.   

Stacking significantly improves sensitivity to GW emission under a
given model.  However, improving detection probability depends upon
stacking according to GW emission models that correctly describe nature.   The storm light
curve motivated two stacking models: a flat-weighted model which includes
the 11 most energetic EM bursts with unity weighting factors; and an EM-fluence-weighted model comprised of the
18 most energetic EM bursts.  The $N=11$ cutoff in the flat
model is motivated by a clear separation in
EM fluence of the 11 most energetic bursts in the storm.
Including the 18 most energetic bursts in the fluence-weighted model
accounts for 95\% of the total EM fluence of the more than 40 bursts in the light
curve.   In the fluence-weighted model, time-frequency excess power tilings
are weighted according to burst-integrated BAT counts before stacking.
Further details are in\,\cite{stackMethod}.

To obtain estimates of the times of EM bursts in the storm,  we measure the intersections of the rapid rising edges 
of each burst with the light curve noise floor measured in a quiescent period of data in the 50\,s BAT sequence before the start of the storm (Fig. \,\ref{fig:lc_rises}).    We correct these times for satellite-to-geocenter 
times-of-flight using the known SGR 1900+14 sky position and Swift ephemeris, which vary from (17.12--17.48)\,ms over the $\sim$30\,s duration of the storm.   The stack-a-flare analysis method is robust to relative timing errors smaller than GW signal durations\,\citep{stackMethod}.
EM fluences are estimated by integrating detector counts under each burst in the light curve.     We conservatively converted counts to fluences using the lower bound of the Konus-Wind total fluence range given above.

We divide the GW data into an on-source time region, in which GWs associated 
with the storm could be expected, and a background region, with statistically similar noise in which we do not expect a GW.  This is done after applying category 1 and category 2 data quality cuts described in\,\cite{s5allsky}.  The on-source region consists of 4\,s of stacked data.  Each 4\,s region comprising the stack is centered on the time of one of the EM bursts included in the GW emission stacking model.    Background regions consist of 1000\,s 
of data on either side of the storm. On-source and background segments are analyzed and stacked identically, and the stacked time-frequency tilings are passed through a clustering algorithm resulting in lists of 
``analysis events.''  Background analysis events due to fluctuating detector noise are used to estimate the significance of on-source events; 
significant events, if any, are subject to vetoes\,\citep{s5allsky}.

Using $\pm$2\,s regions around bursts in the storm accounts for uncertainties in the EM burst times and a possible systematic delay between GW and EM emission.   Although GW emission in SGRs is expected to occur almost 
simultaneously with the EM burst\,\citep{ioka01}, a common bias in trigger times shared by all bursts in the stacking set of $\lesssim$1\,s can be handled with a $\pm$2\,s on-source region.

As in\,\cite{s5y1sgr}, 
this search targets neutron star fundamental mode ringdowns (RDs) predicted in\,\cite{andersson97, pacheco98, ioka01, 
andersson02} as well as short-duration GW signals of unknown waveform.  RDs are targeted because $f$-modes are the most
efficient GW emitters\,\citep{andersson97}.  We assume that given a neutron star, $f$-mode frequencies and damping 
timescales would be similar from event to event, and that unknown signals would at least have similar central frequencies and durations from event to event.

As in\,
\cite{s5y1sgr}, we thus focus on two distinct regions in the target signal time-frequency parameter space.  The first region targets  $\sim$100-400\,ms 
duration signals in the (1--3)\,kHz band, which includes $f$-mode RD signals  predicted in\,\cite{benhar04} for ten realistic neutron star 
equations of state.  We choose a search band of (1--3)\,kHz for RD 
searches, with a 250\,ms time window which was found to give optimal search sensitivity\,\citep{kalmus08}.  The second region targets $\sim$(5--200)\,ms duration signals in the (100--1000)\,Hz band. The target durations are set by prompt SGR burst timescales (5\,ms to 200\,ms) and the target frequencies are set by the detector's sensitive region.  We search 
in two bands: (100--200)\,Hz (probing the region in which the detectors are most sensitive) and (100--1000)\,Hz (for full spectral coverage below 
the RD search band) using a 125\,ms time window.  In all, we search in three frequency bands and two GW emission models (flat and fluence-weighted).  This amounts to a total of six 4\,s-long stacked on-source regions.

We estimate loudest-event upper limits\,\citep{brady04} on GW root-sum-squared 
strain $\hrss$ incident at the detector.  We can 
construct simulations of impinging GW with a given $\hrss$.  Following \cite{s2burst}
 \be 
 h_{\mathrm{rss}}^{2} = h_{\mathrm{rss+}}^{2} + h_{\mathrm{rss\times}}^{2}, 
 \ee
 where e.g.
 \be
 h_{\mathrm{rss+}}^{2} = \int_{-\infty}^{\infty} \abs{h_{+}}^{2} dt 
 \ee
and $h_{\mathrm{+,\times}}(t)$ are the two GW polarizations. The relationship between the GW polarizations and the detector response $h(t)$ 
to an impinging GW from a polar angle and azimuth $(\theta,\phi)$ and with polarization angle $\psi$ is:
 \be
 h(t) = F_{+}(\theta, \phi, \psi) h_+(t)   +  F_{\times}(\theta, \phi, \psi)
 h_{\times}(t)
 \label{eq:hsim}
 \ee
where $F_{+}(\theta, \phi, \psi)$ and $F_{\times}(\theta, \phi, \psi)$ are the antenna functions for the source at $(\theta,\phi)$\,\citep{300years}.   
At the time of the storm, the polarization-independent RMS antenna response $(F_{+}^2 + F_{\times}^2)^{1/2}$, which indicates the average sensitivity to a given sky location, was 0.39 for LIGO Hanford observatory and 0.46 for the LIGO Livingston observatory.

We can also set upper limits on the emitted isotropic GW emission energy $\egw$ at a source distance $R$ associated with $h_{+}(t)$ and ${h}_{\times}(t)$ via\,\citep{shapiro83}
 \be 
 E_{\mathrm{GW}} = 4\pi R^2 \frac{c^3}{16 \pi G} \int_{-\infty}^{\infty}\left((\dot{h}_{+})^2 + (\dot{h}_{\times})^2\right) dt. 
 \ee 

The procedure for estimating loudest-event upper limits in the individual burst search is detailed in\,\cite{kalmus08, s5y1sgr}.   In brief, the 
upper limit is computed in a frequentist framework by injecting artificial signals into the background data 
and recovering them with the search pipeline\,(see for example \cite{S2inspiral,S2S3S4GRB}).   An analysis event is associated with each injection, and compared to the loudest on-source 
analysis event.  The GW strain or isotropic energy at 90\% detection efficiency is the strain or isotropic energy at which 90\% of injections have associated events louder than the loudest on-source event.  

We use the twelve waveform types described in\,\cite{s5y1sgr} to establish detector sensitivity and thereby set upper limits:  linearly and circularly 
polarized RDs with $\tau=200$\,ms and frequencies in the range (1--3)\,kHz; and band- and time-limited white noise bursts (WNBs) with 
durations of 11\,ms and 100\,ms and frequency bands matched to the two lower frequency search bands.  

These waveforms are used to construct compound injections determined by the emission model.  In the flat model,  11 GW bursts comprise a compound injection, each is identical, and our stated $\hrss$ and $\egw$  are for one such GW burst in the compound injection.   In the fluence-weighted model, 18 GW bursts comprise a compound injection, they are weighted (in amplitude) with the square root of integrated counts, and our stated $\hrss$ and $\egw$ are for the loudest GW burst in the compound injection.  A single polarization angle is chosen randomly for every compound injection.  In assuming that the bursts emitted are identical up to an amplitude scale factor, we implicitly assume the star's GW emission mechanism and symmetry axis are constant over bursts in the storm.

\section{Results}

\begin{table*}
\caption[ ]{Stack-a-flare SGR 1900+14 storm upper limits.}
\begin{tabular}{@{\extracolsep{\fill}}l|rlrcc|rlrcc}
 \hline \hline
 & \multicolumn{5}{c}{N=11 Flat} & \multicolumn{5}{c}{N=18 Fluence-weighted} \\ 
 Simulation type &  \multicolumn{3}{c}{$ \hrssn [ 10^{-22} ~ \rthz $] }  & $\egwn$ [erg] & $\gamman$ & \multicolumn{3}{c}{$ \hrssn [ 10^{-22} ~ \rthz $] }  & $\egwn$ [erg] & $\gamman$ \\ 
 \hline 
 WNB 11ms 100-200 Hz   & 1.3 & $^{ +0.0 ~ +0.17 ~ +0.0}$ &  $= 1.5$ & $\sci{1.9}{45}$ & $\sci{3}{4}$ & 2.1 & $^{ +0.0 ~ +0.27 ~ +0.094}$ &  $= 2.4$ & $\sci{5.0}{45}$ & $\sci{3}{4}$ \\
 WNB 100ms 100-200 Hz   & 1.5 & $^{ +0.0 ~ +0.19 ~ +0.0}$ &  $= 1.7$ & $\sci{2.4}{45}$ & $\sci{4}{4}$ & 2.3 & $^{ +0.0 ~ +0.30 ~ +0.098}$ &  $= 2.6$ & $\sci{6.0}{45}$ & $\sci{3}{4}$ \\
 WNB 11ms 100-1000 Hz   & 3.5 & $^{ +0.0 ~ +0.45 ~ +0.0}$ &  $= 3.9$ & $\sci{1.8}{47}$ & $\sci{3}{6}$ & 5.2 & $^{ +0.0 ~ +0.67 ~ +0.29}$ &  $= 5.9$ & $\sci{4.1}{47}$ & $\sci{2}{6}$ \\
 WNB 100ms 100-1000 Hz   & 3.8 & $^{ +0.0 ~ +0.50 ~ +0.0}$ &  $= 4.3$ & $\sci{2.0}{47}$ & $\sci{3}{6}$ & 5.6 & $^{ +0.0 ~ +0.73 ~ +0.29}$ &  $= 6.3$ & $\sci{4.5}{47}$ & $\sci{2}{6}$ \\
 \hline
 RDC 200ms 1090 Hz   & 4.5 & $^{ +0.045 ~ +0.59 ~ +0.0}$ &  $= 5.2$ & $\sci{1.2}{48}$ & $\sci{2}{7}$ & 7.2 & $^{ +0.072 ~ +0.93 ~ +0.33}$ &  $= 8.2$ & $\sci{3.0}{48}$ & $\sci{2}{7}$ \\
 RDC 200ms 1590 Hz   & 6.4 & $^{ +0.19 ~ +0.84 ~ +0.0}$ &  $= 7.4$ & $\sci{5.1}{48}$ & $\sci{8}{7}$ & 11 & $^{ +0.33 ~ +1.4 ~ +0.44}$ &  $= 13$ & $\sci{1.5}{49}$ & $\sci{8}{7}$ \\
 RDC 200ms 2090 Hz   & 9.3 & $^{ +0.28 ~ +1.8 ~ +0.41}$ &  $= 11$ & $\sci{2.1}{49}$ & $\sci{3}{8}$ & 14 & $^{ +0.43 ~ +2.8 ~ +0.72}$ &  $= 18$ & $\sci{4.9}{49}$ & $\sci{3}{8}$ \\
 RDC 200ms 2590 Hz   & 11 & $^{ +0.34 ~ +2.2 ~ +0.32}$ &  $= 14$ & $\sci{4.6}{49}$ & $\sci{8}{8}$ & 17 & $^{ +0.50 ~ +3.3 ~ +1.0}$ &  $= 21$ & $\sci{1.0}{50}$ & $\sci{5}{8}$ \\
 \hline
 RDL 200ms 1090 Hz   & 9.3 & $^{ +0.0 ~ +1.2 ~ +0.95}$ &  $= 11$ & $\sci{5.3}{48}$ & $\sci{9}{7}$ & 16 & $^{ +0.0 ~ +2.1 ~ +1.6}$ &  $= 18$ & $\sci{1.5}{49}$ & $\sci{8}{7}$ \\
 RDL 200ms 1590 Hz   & 14 & $^{ +0.42 ~ +1.8 ~ +1.1}$ &  $= 17$ & $\sci{2.6}{49}$ & $\sci{4}{8}$ & 19 & $^{ +0.58 ~ +2.5 ~ +1.9}$ &  $= 23$ & $\sci{5.1}{49}$ & $\sci{3}{8}$ \\
 RDL 200ms 2090 Hz   & 20 & $^{ +1.2 ~ +3.9 ~ +1.4}$ &  $= 25$ & $\sci{1.0}{50}$ & $\sci{2}{9}$ & 27 & $^{ +1.6 ~ +5.3 ~ +2.8}$ &  $= 34$ & $\sci{1.9}{50}$ & $\sci{1}{9}$ \\
 RDL 200ms 2590 Hz   & 25 & $^{ +1.8 ~ +5.0 ~ +3.0}$ &  $= 33$ & $\sci{2.6}{50}$ & $\sci{4}{9}$ & 39 & $^{ +2.7 ~ +7.7 ~ +2.5}$ &  $= 50$ & $\sci{6.2}{50}$ & $\sci{3}{9}$ \\
 \hline \label{table:bestresults} 
 \end{tabular} 
 \vspace{0.12in}
 \end{table*} 

We find no statistically significant GW signal associated with the SGR
1900+14 storm.
The significance of on-source analysis events is inferred by noting the rate at which background analysis events of equal or greater loudness occur.  We examined 4\,s stacked on-source regions in the flat and fluence-weighted models in the three search bands.  The most significant 
on-source analysis event from these six searches was from the flat model in the (100--1000)\,Hz band and had a corresponding background rate of $\sci{5.0}{-2}$\,Hz (1 per 20\,s) in that search.  

\begin{figure}[!t]
\begin{center}
\includegraphics[angle=0,width=90mm, clip=false]{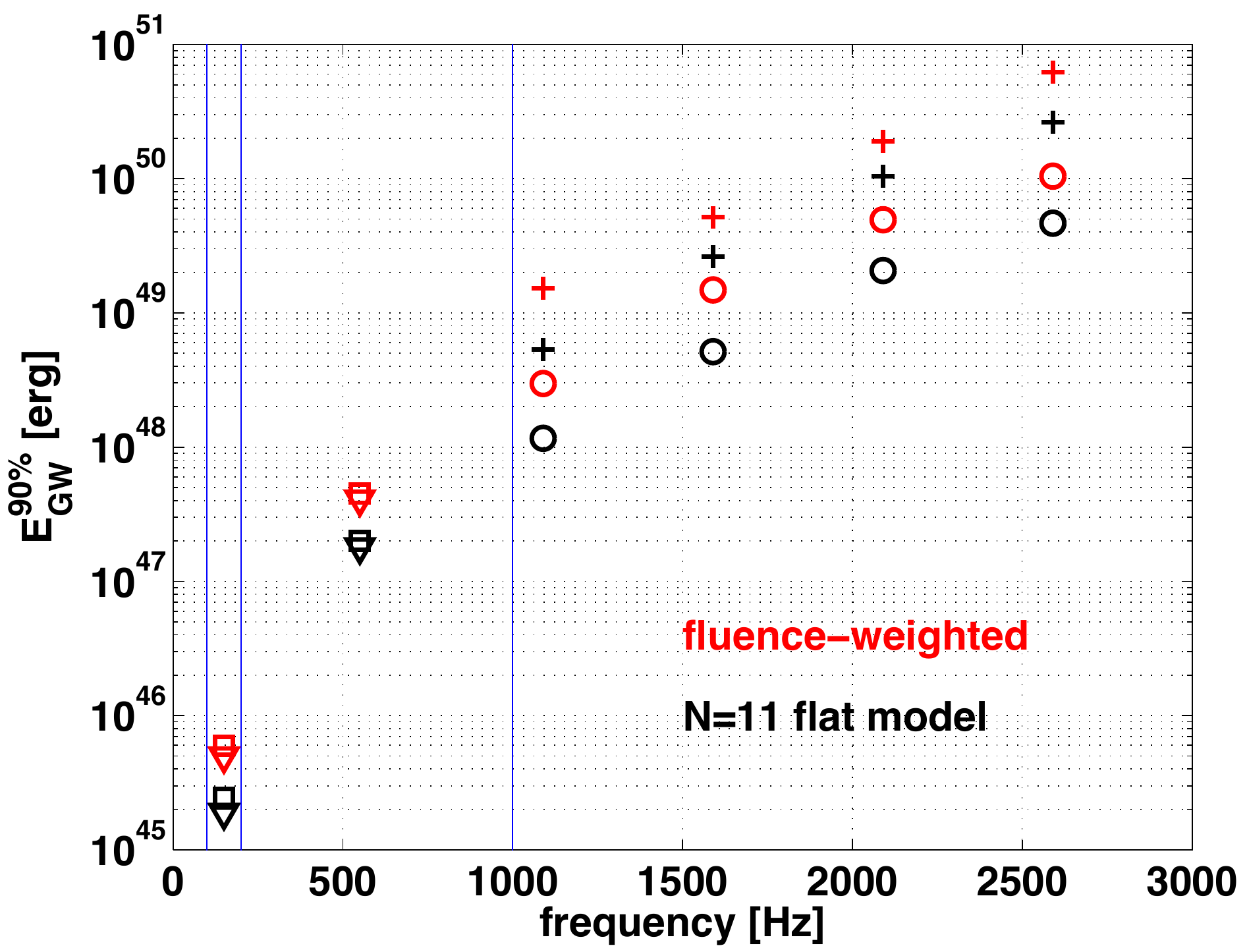}
\caption[Stack-a-flare SGR 1900+14 storm upper limit estimates] {Stack-a-flare SGR 1900+14 storm isotropic energy upper limit estimates at 10\,kpc, for flat and fluence-weighted
emission models.   We set upper limits at characteristic points in the signal parameter space in order to quantify the meaning of our non-detection result.  Uncertainties have been folded in.  Vertical lines indicate boundaries of the three distinct search frequency bands.  Crosses and circles indicate linearly and circularly polarized RDs, respectively.  Triangles and squares represent 11\,ms and 100\,ms band- and time-limited WNBs, 
respectively.  Symbols are placed at the waveform central frequency.  These results reflect the noise curves of the detectors.  }
\label{fig:bestresultsegw}
\end{center}
\end{figure}

Table\,\ref{table:bestresults} and Fig. \,\ref{fig:bestresultsegw} give model-dependent loudest-event upper limits at 90\% detection efficiency computed for the GW signal associated with the single loudest EM burst.  We give strain upper limits ($\hrssn$) and isotropic emission energy upper limits  at a nominal SGR 1900+14 distance of 10\,kpc ($\egwn$).  We also give upper limits $\gamman =\egwn/E_{\mathrm{EM}}$, a source-distance-independent measure of the extent to which an energy upper limit probes the GW emission efficiency, calculated using a conservative estimate of $\sci{1.0}{-4}$\,erg cm$^{-2}$ for the total fluence of the storm to estimate fluences for individual peaks.   In the fluence-weighted model, $\gamma$ is the same for each individual burst.  In the flat model we report the mean value of $\gamma$ for the 11 bursts.

Superscripts in Table\,\ref{table:bestresults} give a systematic error and uncertainties at 90\% confidence. (Similar estimates were made for the $\egwn$ but are not shown in the table.) The first and second superscripts account for the systematic error and statistical uncertainty, respectively, in the detector calibrations. The third is the statistical uncertainty from using a finite number of trials (200) in the Monte Carlos, estimated with the bootstrap method using 200 ensembles\,\citep{efron79}.  The systematic error and the quadrature sum of the statistical uncertainties are added to the final sensitivity estimates.   One-sigma burst timing uncertainties from fits of burst rising edges are accounted for in the Monte Carlo simulations.  Estimating uncertainties is further described in\,\cite{stackMethod}.  

\section{Discussion}  

The stacked search described here extends the recent LIGO search for GW associated with the 2004 SGR 
1806--20 giant flare and 190 lesser events from SGR 1806--20 and SGR 1900+14\,\citep{s5y1sgr}.   That search was the first search sensitive to neutron star $f
$-modes, and it set individual burst upper limits  $\egwn$ ranging from $\sci{3}{45}$\,erg to $\sci{9}{52}$\,erg (depending on waveform type and detector antenna factors and noise 
characteristics at the time of the burst), but did not detect any GWs.  The best values of $\gamman$ in\,
\cite{s5y1sgr}, for the giant flare, were in the range $\sci{5}{1}$--$\sci{6}{6}$ depending on waveform type.

The upper limits obtained here are a factor of 12 more sensitive in energy
than the SGR 1900+14 storm upper limits in \,\cite{s5y1sgr}, which analyzed the storm in a single
$\pm20$\,s on-source region. Those previous limits already overlapped the range of EM
energies seen in the loudest flares as well as the range of GW energies predicted by the most extreme models~\citep{ioka01}.  The flat model gives isotropic energy upper limits on average a factor of 4 lower than a reference $N=1$ (non-stacked) scenario (with a $\pm2$\,s on-source region) and a factor of 2 lower than the fluence-weighted model.  
However, our storm $\gamma $ upper limits are still a few hundred
times the SGR 1806--20 giant flare $\gamma$ upper limits, due to the
tremendous EM energy released by the giant flare.
There is very little discussion of $\gamma$ in the theory literature with which to
compare.  

The Advanced LIGO detectors promise an improvement in energy sensitivity of
more than a factor of 100.
Furthermore, on 2008 August 22, SGR 0501+4516 was
discovered\,\citep{gcn8112, gcn8113, gcn8115} and may be located only
1.5\,kpc away\, \citep{gcn8149,
leahy95}.  SGR 0501+4516 searches will thus gain an additional 2 orders of magnitude in energy and $\gamma$ upper limits compared 
to SGRs at 10\,kpc.  A stacking analysis of SGR 0501+4516 bursts with 
Advanced LIGO (a gain of 4 orders of magnitude in energy sensitivity) could therefore reach $\gamma$ values below unity, even 
without another giant flare.

In the future we plan to carry out stacking searches on isolated SGR bursts, and eventually to perform searches using Advanced
LIGO data.
Our stacked upper limits depend on theoretical guidance as to what
weightings and time delays are possible, and the significance of our
results depends on predictions of the range of $\egw$ and $\gamma$; yet all
of these things are scarce.
We hope that our continued efforts to search for GW associated with SGR and
Anomalous X-ray Pulsar bursts encourage further modeling of GW emission from these intriguing
objects.

\acknowledgments
The authors are grateful to the Swift team for the SGR 1900+14 storm data.
The authors gratefully acknowledge the support of the United States National
Science Foundation for the construction and operation of the LIGO
Laboratory and the Science and Technology Facilities Council of the
United Kingdom, the Max-Planck-Society, and the State of
Niedersachsen/Germany for support of the construction and operation
of the GEO600 detector. The authors also gratefully acknowledge the
support of the research by these agencies and by the Australian
Research Council, the Council of Scientific and Industrial Research
of India, the Istituto Nazionale di Fisica Nucleare of Italy, the
Spanish Ministerio de Educaci\'{o}n y Ciencia, the Conselleria
d'Economia Hisenda i Innovaci\'{o} of the Govern de les Illes
Balears, the Royal Society, the Scottish Funding Council, the
Scottish Universities Physics Alliance, The National Aeronautics and
Space Administration, the Carnegie Trust, the Leverhulme Trust, the
David and Lucile Packard Foundation, the Research Corporation, and
the Alfred P. Sloan Foundation. This Letter is LIGO-\ligodoc.

\end{document}